\documentclass[draftclsnofoot,onecolumn,doublespace,12pt,letter]{IEEEtran}
\ifCLASSINFOpdf
\else
\fi
\usepackage{cite}
\usepackage{enumerate}
\usepackage[center]{caption}
\usepackage{amssymb}
\usepackage{graphicx}
\usepackage[cmex10]{amsmath}

\usepackage{algorithmic}
\usepackage{url}

\usepackage{algorithm}

\usepackage{latexsym}

\usepackage{slashed}
\usepackage{centernot}
\newcommand{\bs}{\boldsymbol}
\newcommand{\bt}{\textbf}
\DeclareMathOperator{\sprod}{\bs{\centernot\prod}}

\begin{document}
%
\title{Turbo Receiver Design for Phase Noise Mitigation in OFDM Systems}



%
\author{\IEEEauthorblockN{Gokul Sridharan\IEEEauthorrefmark{0}
 and Teng Joon Lim\IEEEauthorrefmark{0}
}
\IEEEauthorblockA{\IEEEauthorrefmark{0}Electrical and Computer Engineering Department\\
University of Toronto, Canada\\
Email:gsridharan@comm.utoronto.ca, limtj@comm.utoronto.ca}
}


\maketitle

\begin{abstract}
This paper addresses the issue of phase noise in OFDM systems. Phase noise (PHN) is a transceiver impairment resulting from the non-idealities of the local oscillator. We present a case for designing a turbo receiver for systems corrupted by phase noise by taking a closer look at the effects of the common phase error (CPE). Using an approximate probabilistic framework called variational inference (VI), we develop a soft-in soft-out (SISO) algorithm that generates posterior bit-level soft estimates while taking into account the effect of phase noise. The algorithm also provides an estimate of the phase noise sequence. Using this SISO algorithm, a turbo receiver is designed by passing soft information between the SISO detector and an outer forward error correcting (FEC) decoder that uses a soft decoding algorithm. It is shown that the turbo receiver achieves close to optimal performance. 
\end{abstract}


%
\IEEEpeerreviewmaketitle

\section{Introduction}
\label{section_intro}
Increase in the demand for higher data rates has led to OFDM becoming the technology of choice for next generation wireless standards such as WiMAX and LTE. The deployment of the devices built around these standards brings to light the various implementation issues that become critical when designing an OFDM based system. PHN is an impairment that is quite different from the other impairments since it is a continuous time noise process that cannot be compensated for in the training stage. In this paper we consider the problem of data detection in an OFDM system that has been corrupted by PHN.\\

PHN arises from imperfections in the local oscillator (LO) that result in a random drift of the LO phase from its reference and is commonly modelled as a Wiener process. But, in general, a local oscillator is usually followed by a phase locked loop (PLL) that keeps the phase and frequency matched to a reference and in such a scenario PHN is better modelled as a wide sense stationary (WSS) process with bounded variance \cite{petrovic2}. In this paper we assume phase noise to be a first order auto-regressive (AR(1)) process as suggested in \cite{Webster} for the IEEE 802.11g standard. The effect of PHN has been studied extensively \cite{pollet1,petrovic1,wu1,wu2}. The effect of PHN can be split into two: the rotation of all the sub-carriers by a certain angle called the common phase error (CPE) and the leakage of the neighboring sub-carriers resulting in ICI. CPE is the average of the PHN sequence spanning an OFDM symbol.

 PHN mitigation for single carrier systems and multi carrier systems such as OFDM has been studied extensively \cite{wu1},\cite{darryl1,septier1,gong1,rave1,barbieri1}. While for single carrier systems it is possible to construct the factor graph and adopt a message passing algorithm \cite{barbieri1} such an approach becomes prohibitively complicated for OFDM systems. Most algorithms suggested for PHN estimation in OFDM systems involve a decision directed process where the data symbols are estimated ignoring PHN and these decisions are fed back to estimate the PHN. This process is iterated either for a fixed number of times or until convergence. These are hard decision algorithms and do not provide any soft information on the decisions made. Another class of techniques that have been used to jointly estimate PHN and data symbols include the Markov chain Monte Carlo methods \cite{septier1}. The major concern with decision directed algorithms is the assumption that the initial crude estimation of data symbols ignoring PHN results in a majority of the symbols being detected correctly while a good fraction of the remaining symbols get corrected over the subsequent iterations. As we will see in a subsequent section, this need not necessarily be true and can lead to a premature error floor. This motivates the need for a turbo receiver that exchanges messages between the symbol detector and the outer FEC decoder.

While \cite{darryl1} developed algorithms for PHN mitigation using VI that generated soft symbol estimates using a Gaussian approximation, we design a soft detection algorithm that computes soft bit estimates of the transmitted bits using the VI framework and a discrete distribution for the bits. The algorithm is capable of incorporating prior information on bits and hence fits in naturally with a soft decoding algorithm for the FEC code. The paper is organized as follows. Section \ref{section_PHN} reviews modelling of phase noise, section \ref{section_signal_model} sets up the received signal model and discusses the consequences of phase noise on the received signal. In section \ref{section_algorithm} the VI framework for this problem is presented along with the bit level detection algorithm. The last section presents the simulation results.

\section{Phase Noise Characteristics}
\label{section_PHN}
Since almost all communication systems employ an LO followed by a PLL, we assume PHN to be a WSS process. In particular, we model it as an AR(1) process. Such a model has been adopted by IEEE standards such as 802.11g where phase noise is modelled as the output of a single pole Butterworth filter driven by a zero mean white noise process. Such a process has an autocorrelation function of the form \cite{darryl1}
\begin{equation}
 R_p(k)=\sigma_\theta^2e^{-2\pi\Omega_oT_s|k|}.
\end{equation}
Here, $1/T_s$ is the sampling rate and $\Omega_o$ is the one sided 3-dB bandwidth of the oscillator. $\sigma_\theta$ is the RMS value of the PHN process in radians, a typical value being around 3$^\circ$.  In general, the 3-dB bandwidth of an oscillator tends to be in the range of hundreds of kHz. One can write the covariance matrix of a length-N sequence of PHN as 
\begin{equation}
\label{phase_cov_matrix}
\boldsymbol \Phi=\sigma^2_\theta \begin{bmatrix}
          1 & p & p^2 & \hdots & p^{N-1} \\
	  p & 1 & p^3 & \hdots & p^{N-2}\\
	  \vdots & \vdots & \ddots & \hdots \\
	 p^{N-1} & p^{N-2} & p^{N-3} &\hdots & 1 \\	
         \end{bmatrix}
\end{equation}
where, $p=e^{-2\pi\Omega_oT_s}$. Usually, the cut off frequency of the Butterworth filter ($\Omega_o$) is set to the 3-dB bandwidth of the oscillator.

\subsection{Sample mean statistics}
Since the CPE plays a critical role in the detection of symbols in affected by PHN, we take a closer look at the sample mean of a length-N PHN sequence. Suppose we let $\bar{\theta}$ denote the sample mean of a length-N sequence of the phase noise process, we have,
\begin{equation}
\bar{\theta}=\frac{1}{N} \sum_{k=1}^N\theta[k].
\end{equation}
Since the statistics of the phase noise process are assumed to be known, it can be shown that $\bar{\theta}$ is a zero mean Gaussian random variable with variance $\bt 1^T\bs\Phi\bt 1/N^2$ \cite{darryl3}. Note that the variance of the sample mean is a function of N and goes to zero as $N\rightarrow \infty$.

\section{Received Signal Model}
\label{section_signal_model}
\subsection{Gray mapping : bits to M-QAM symbols ( from \cite{darryl2} )}
We assume M to be a perfect square. Let L denote $\log_2$M and let \textbf{B} $\in$ $\{-1,1\}^{\text{N}\times\text{L}}$ represent a matrix of N$\times$L bits that need to be mapped to N M-QAM symbols. Denote the columns of \bt{B} as $\bt{b}_{r1}$, $\bt{b}_{r2}$, $\hdots$ $\bt{b}_{rL/2}$,$\bt{b}_{i1}$, $\bt{b}_{i2}$ $\hdots$ $\bt{b}_{iL/2}$. If the resulting vector of M-QAM symbols is represented as \textbf{d}, then, from \cite{darryl2},  we have the following relation :\\ 
\begin{equation}
 \bt{d}=\sum_{l=1}^{L/2}2^{l-1}\sprod\displaylimits_{p=1}^{p=L/2}\bt{b}_p+j\sum_{l=L/2+1}^{L}2^{l-1}\sprod\displaylimits_{p=L/2+1}^{p=L}\bt{b}_p
\end{equation}

Here, $\sprod$ represents element wise product of the vectors. Let the function \textit{f} denote the mapping from bits to symbols. Thus, $\bt{d}=\textrm{\textit{f}}(\bt{B})$. Further, denote the matrix [$\bt{b}_{r1}$ $\bt{b}_{r2}$ $\hdots$ $\bt{b}_{rL/2}$] as $\bt{B}_r$, and the matrix [$\bt{b}_{i1}$ $\bt{b}_{i2}$ $\hdots$ $\bt{b}_{iL/2}$] as $\bt{B}_i$ and define the functions \textit{f}$_r$ and \textit{f}$_i$ as :
\begin{align}
\textrm{\textit{f}}_i(\bt{B}_i)=j.\Im({\textrm{\textit{f}}(\bt{B})})\\
\textrm{\textit{f}}_r(\bt{B}_r)=\Re({\textrm{\textit{f}}(\bt{B})})
\end{align}
Hence, $\bt d=\textrm{\textit{f}}(\bt{B})=\textrm{\textit{f}}_r(\bt{B}_r) + \textrm{\textit{f}}_i(\bt{B}_i)$.

\subsection{The received signal}
In this work, we consider the detection of an OFDM symbol transmitted over a block fading frequency selective channel, where the channel stays constant over the duration of one OFDM symbol. We also assume that perfect frame synchronization, including carrier frequency recovery have been established in the training stage. We further assume that current channel conditions have been estimated during the training phase and that channel state information is available on the receiver side. Algorithms that can estimate the channel in the presence of PHN and carrier frequency offset have been presented in \cite{septier1},\cite{darryl3}. In the data detection stage we assume that the received symbol vector has been affected by PHN in addition to the channel and the additive noise.  The received signal for such a scenario in the discrete domain after appropriate sampling and removal of the cyclic prefix is given by \\
\begin{equation}
\textbf{r}=\textbf{PF}^H\textbf{Hd}+\textbf{n}=\textbf{PF}^H\textbf{H}\textrm{\textit{f}}(\bt{B})+\textbf{n}.
\end{equation}
Here, \textbf{F} is an $N\times N$ DFT matrix with the (l, m)th entry given by $\bt F_{lm}=(1/\sqrt{N})e^{-(2\pi j(l-1)(m-1)/N)}$, \textbf{P} is the diagonal matrix given by diag($ e^{ j\bs \theta}$) $\approx$ diag($\textbf{1}+j\boldsymbol{\theta}$), where $\boldsymbol{\theta}$ is the PHN sequence and $\textbf{H}=\text{diag}(\bt h)$ is the channel matrix in the frequency domain and the $N\times L$ binary matrix \textbf{B} contains the transmitted bit sequence (\bt d is the corresponding symbol sequence). \bt n is complex white Gaussian noise with variance $\sigma^2$ per dimension.

Detection schemes that ignore PHN involve computing the DFT of the received signal and then adopting MMSE or zero forcing detection on individual sub carriers. The DFT of the received signal in the presence of phase noise results in a vector $\bt R=[R_0 R_1 \hdots R_{N-1}]^T$ which can be written as \cite{darryl1} 
\begin{align}
\label{ici_eq}
R_k=c_0d_kh_k+\sum_{l=0,l\neq k}^{N-1}d_l h_lc_{(l-k)mod\: N} + \nu_k.
\end{align}

Here, the vector $\bt c=[c_0 c_1 \hdots c_{N-1}]^T$ is given by $(1/\sqrt{N})\bt {Fp}$ (where $p=e^{j\bs \theta}$) i.e. the frequency domain representation of the PHN sequence. It can be shown that $\bs\nu$ is an uncorrelated white noise process with $ \nu_k \sim \mathcal{CN}(0, 2\sigma^2)$. Eq.(\ref{ici_eq}) clearly illustrates how PHN affects the received signal. Note that $c_0$ is the CPE i.e $1+\frac{1}{N}\sum \theta_k$ (under small angle assumption), and its effect is to rotate every received symbol by the average phase angle $\bar{\theta}$. The next subsection further elaborates the consequences. 

\subsection{Effect of phase noise on the received signal and its consequences}

Decision directed feedback mechanisms that have been suggested to counter the ICI resulting from PHN involve an initial step where the symbols are detected in a crude manner, ignoring the effects of PHN.
The underlying assumption is that a majority of symbols detected in this way are correct and hence a back substitution of these symbols into the received signal structure must aid the detection process. This need not be true for certain realizations of the phase noise, specifically, scenarios where the CPE is high. This is because the received symbol gets rotated by an angle equal to the CPE and if this angle is greater than the average angular separation between the adjacent symbols of the constellation, the symbol is likely to be detected in error. As an example, for the 64-QAM constellation a rotation of 9$^\circ$ causes a symbol error with probability 0.43 in the absence of noise. When the oscillator linewidth is set to 100kHz and the RMS value of phase noise is set to $3^\circ$, the probability of encountering a length 64 (assume 64 sub-carriers in one OFDM symbol) phase noise sequence whose mean is greater than $9^\circ$ is $4\times10^{-5}$. This leads to a premature error floor.\\

One can counter this by either embedding pilots in the OFDM symbol and using them to estimate and compensate for CPE \cite{wu1} or by designing a blind turbo receiver. While embedding pilots is a common practice to aid channel estimation, this comes at the cost of spectral efficiency. In scenarios where embedding pilots in every OFDM symbol is not an option one needs to consider designing a turbo receiver. Since the variance of the CPE decreases with increasing length of the PHN sequence, there exists a compelling case to design a turbo receiver for short length OFDM symbols. Having made the case for joint detection-decoding, we design a bit level detection algorithm that generates soft bit estimates and use it  to set up a turbo receiver. The next subsection explains the basic principles involved.



\subsection{Actual and postulated posterior distributions}

From \cite{darryl1}, we note that by approximating $e^{j\bs\theta}$ to $\bt 1+j\bs\theta$ under the small angle assumption, one can write the conditional distribution of \bt r given \bt B and $\bs\theta$ to be :

\begin{equation}
\label{cond_eq1}
p(\textbf{r}|\textbf{B},\bs\theta)=\mathcal{CN}\big{(}\text{diag}(\textbf{PF}^H\textbf{H}\textrm{\textit{f}(\bt{B})})(\textbf{1}+j\boldsymbol{\theta}), 2\sigma^2\textbf{I}\big{)}
\end{equation}

Further, in \cite{darryl1} it was shown that the optimal detector (i.e. ML estimate of \bt d) for such a signal model, given the prior distribution of the phase noise sequence $\bs\theta$ has an exponential complexity in N. Given that the distribution $p(\bt r|\bt B)$ and consequently the posterior distribution $p(\bt B|\bt r)$ do not lend themselves to efficient optimal detection, we look at an approximation to $p(\bt B|\bt r)$, such that, computing the optimal estimate corresponding to the approximated distribution is straightforward. To this end, we note that p(\textbf{B}$|$\textbf{r}) is computed by marginalizing p(\textbf{B},$\boldsymbol{\theta}|$\textbf{r}) with respect to $\boldsymbol{\theta}$, i.e.
\begin{equation}
\text{p}(\textbf{B}|\textbf{r})=\int \text{p}(\textbf{B},\boldsymbol{\theta}|\textbf{r}) d\boldsymbol{\theta}
\end{equation}
The variational inference approach approximates p(\textbf{B},$\boldsymbol{\theta}|$\textbf{r}) with a function Q(\textbf{B},$\boldsymbol\theta$) of the form Q$_{\bt B}$(\textbf{B})Q$_{\bs\theta} $($\boldsymbol{\theta}$) such that 
\begin{equation}
 \text{Q}(\textbf{B},\boldsymbol\theta) \sim \text{p}(\textbf{B},\boldsymbol{\theta}|\textbf{r}) 
\end{equation}

This kind of an approximation is equivalent to assuming that \bt B and $\bs\theta$ are independent conditioned on \bt r and this in turn implies the maximizer of the distribution Q$_{\bt B}$(\bt B) is the optimal estimate of \bt B. We further assume that the distribution Q$_{\bt B}$(\bt B) can be factorize into $\prod_{n=1}^{N}\prod_{l=1}^{L}Q_b(b_{nl})$. Such a factorization for the postulated posterior distribution of the bits, where the distribution is assumed to be independent over \textit{n} and \textit{l}, is commonly known as the mean field approximation. In this paper, we assume Q$_b(b_{nl})$ to be a Bernoulli distribution with parameter $\lambda_{nl}$. We postulate the posterior conditional distribution of phase noise to be a multi-variate Gaussian distribution with mean $\bt m_{\bs\theta}$ and covariance $\bt S_{\bs\theta}$. Thus, we have 
\begin{align}
\text{Q}(\textbf{B},\boldsymbol{\theta})&=\text{Q}_{\bt B}(\textbf{B}) \text{Q}_{\bs \theta}(\boldsymbol{\theta})\nonumber \\
=& \left[\prod_{n=1}^{N} \prod_{l=1}^{L}(\lambda_{nl})^{\frac{1+b_{nl}}{2}} (1-\lambda_{nl})^{\tfrac{1-b_{nl}}{2}}\right]\mathcal{N}(\textbf{m}_\theta,\textbf{S}_\theta)\\
=&\left[\prod_{n=1}^{N} \prod_{l=1}^{L}\left (\tfrac{1+\hat{b}_{nl}}{2}\right )^{\frac{1+b_{nl}}{2}} \left (\tfrac{1-\hat{b}_{nl}}{2}\right )^{\tfrac{1-b_{nl}}{2}}\right]  \mathcal{N}(\textbf{m}_\theta,\textbf{S}_\theta)
\end{align}
where, $\hat{b}_{nl}=2\lambda_{nl}-1$, is the mean of the postulated posterior distribution Q$_b(b{_{nl}})$. Having fixed the structure of the postulated posterior distribution, it remains now to compute the parameters of this distribution such that it closely approximates the actual posterior distribution. The variational inference approach introduces the concept of variational free energy as a measure of similarity between two distributions. The variational free energy between the two distributions of interest here is given by 
\begin{equation}
\label{free_energy_eq}
F(Q,p)=\int _{\textbf{B},\boldsymbol{\theta}} \text{Q}(\textbf{B},\boldsymbol{\theta}) \log \frac{\text{Q}(\textbf{B},\boldsymbol\theta)}{\text{p}(\textbf{B},\boldsymbol{\theta},\textbf{r})} d\textbf{B}d\boldsymbol{\theta}.
\end{equation}

We use the distribution $\text{p}(\textbf{B},\boldsymbol{\theta},\textbf{r})$ instead of $\text{p}(\textbf{B},\boldsymbol{\theta}|\textbf{r})$ as they are related simply through a constant of proportionality. It is to be noted that the free energy expression is exactly equal to the Kullback-Leibler divergence between Q$(\bt B,\bs\theta)$ and $\text{p}(\textbf{B},\boldsymbol{\theta}|\textbf{r})$ to within an additive constant. The variational inference approach involves the minimization of the free energy over the parameters of Q($\bt B,\bs\theta$) so that the resulting distribution closely approximates the actual posterior distribution.

The actual posterior distribution can be computed by conditioning over the unknowns as follows:
\begin{align}
\text{p}(\textbf{B},\boldsymbol{\theta},\textbf{r})  =& \text{p}(\textbf{r}|\textbf{B},\boldsymbol{\theta})\text{p}(\textbf{B})\text{p}(\boldsymbol{\theta}) 
\end{align}
Using (\ref{cond_eq1}), and assuming independent Bernoulli distributed priors on bits with the l$^{th}$ bit of the n$^{th}$ symbol having mean $\mu_{nl}$ (set $\mu_{nl}$ to 0 if no prior information is available) and assuming the prior distribution of PHN to be a multi-variate Gaussian distribution with mean $\bs\mu_\theta$ and covariance matrix $\bs\phi_\theta$ (if no prior information is available, set mean to \bt 0 and covariance to $\bs\Phi$  as given in (\ref{phase_cov_matrix})), we can write the posterior distribution as

\begin{align}
\text{p}(\textbf{B},\boldsymbol{\theta},\textbf{r})    =\mathcal{CN}\big{(}\text{diag}(\textbf{PF}^H\textbf{H}\textrm{\textit{f}(\bt{B})})(1+j\boldsymbol{\theta}), 2\sigma^2\textbf{I}\big{)}\hspace{-0.2cm}
\left[\prod_{n=1}^{N} \prod_{l=1}^{L}\left(\tfrac{1+\mu_{nl}}{2}\right)^{\tfrac{1+b_{nl}}{2}} \left(\tfrac{1-\mu_{nl}}{2}\right)^{\tfrac{1-b_{nl}}{2}}\right] 
\mathcal{N}(\boldsymbol\mu_\theta,\boldsymbol\phi_\theta).
\end{align}

The next section discusses the computation and minimization of the free energy expression.
\section{The bit-level Variational Inference Algorithm}
\label{section_algorithm}
\subsection{Free Energy Evaluation}
The free energy expression given in (\ref{free_energy_eq}) can be written as the summation of five terms as shown in (\ref{free_energy_eq2}).
\begin{align}
\label{free_energy_eq2}
F(Q,p)
=& -\underbrace{\int_{\textbf{B}}Q(\boldsymbol\theta)\log p(\textbf{B})d\textbf{B}}_{i}
 -\underbrace{\int_{\boldsymbol\theta}Q(\boldsymbol\theta)\log p(\boldsymbol\theta)d\boldsymbol\theta}_{ii}
+\underbrace{\int_{\textbf{B}}Q(\textbf{B})\log p(\textbf{B})d\textbf{B}}_{iii} \nonumber \\
&+\underbrace{\int_{\boldsymbol\theta}Q(\boldsymbol\theta)\log Q(\boldsymbol\theta)d\boldsymbol\theta}_{iv} 
-\underbrace{\int _{\textbf{B},\boldsymbol\theta}Q(\textbf{B})Q(\boldsymbol\theta)\log(p(\textbf{r}|\textbf{B},\boldsymbol\theta))d\textbf{B}d\boldsymbol\theta}_{v} 
\end{align}

 The exact computation of the free energy expression is given in Appendix \ref{appendix_free_energy_eval} and the final expression is presented here. Note that the free energy expression is parameterized by the mean and the covariance matrix of the PHN sequence and the mean value of the posterior estimate of the bits. Note that when treating the unknown \bt B as a matrix of Bernoulli random variables, we denote $\hat{\bt B}$ as the matrix of the means       \mbox{($\hat{b}_{nl}$)} of the corresponding random variables in \bt B. We define $\hat{\bt B}_r$, $\hat{\bt B}_i$, $\hat{\bt b}_{rj}$ and $\hat{\bt b}_{rk}$ in a similar fashion.

\begin{align}
\label{final_free_eq}
F(\textbf{m}_\theta,\textbf{S}_\theta,\hat{\textbf{B}})=&- \sum _{n=1}^N \sum _{l=1}^{L} \left[ \left(\tfrac{1+\hat{b}_{nl}}{2} \right)\log \left(\tfrac{1+\mu_{nl}}{2} \right)\hspace{-0.1cm}+\hspace{-0.1cm} \left(\tfrac{1-\hat{b}_{nl}}{2} \right)\log \left(\tfrac{1-\mu_{nl}}{2} \right) \right] \hspace{-0.1cm} 
 +\hspace{-0.08cm}\frac{1}{2}\text{tr}(\boldsymbol{\phi}^{-1}_\theta \textbf{S}_\theta) \hspace{-0.08cm} +\hspace{-0.08cm}\frac{1}{2}\textbf{m}_\theta^T\boldsymbol\phi^{-1}_\theta \textbf{m}_\theta \nonumber \\
 &- \boldsymbol\mu_\theta^T \boldsymbol\phi^{-1}_{\theta}\textbf{m}_\theta 
+ \sum_{n=1}^N \sum_{l=1}^L \left[ \left(\tfrac{1+\hat{b}_{nl}}{2} \right) \log \left(\tfrac{1+\hat{b}_{nl}}{2} \right)+ \left( \tfrac{1-\hat{b}_{nl}}{2} \right) \log \left(\tfrac{1-\hat{b}_{nl}}{2} \right) \right] 
-\frac{1}{2}\log|\textbf{S}_\theta| \nonumber \\
&+\frac{1}{2\sigma^2}\Big[ -\textbf{r}^H\textbf{Z}\textrm{\textit{f}($\hat{\bt{B}}$)}-\textrm{\textit{f}($\hat{\bt{B}}$)}^H\textbf{Z}^H\textbf{r} 
 +\textrm{\textit{f}$_r$($\hat{\bt{B}}_r$)}^H \bt M_0\textrm{\textit{f}$_i$($\hat{\bt{B}}_i$)} 
+ \textrm{\textit{f}$_i$($\hat{\bt{B}}_i$)}^H \bt M_0\textrm{\textit{f}$_r$($\hat{\bt{B}}_r$)} \nonumber \\
&+\textrm{\textit{f}$_r$($\hat{\bt{B}}_r$)}^H \bt M_1\textrm{\textit{f}$_r$($\hat{\bt{B}}_r$)} 
+\textrm{\textit{f}$_i$($\hat{\bt{B}}_i$)}^H \bt M_1\textrm{\textit{f}$_i$($\hat{\bt{B}}_i$)}
+ \textbf{1}^H \bt M_2\bs{\nu}_r + \textbf{1}^H \bt M_2\bs{\nu}_i \Big]  
\end{align}

\subsection{Free Energy Minimization}
Clearly, closed form expressions for the optimal parameters that minimize the free energy expression cannot be computed. Hence, we adopt a coordinate-descent approach to the minimization, wherein, one parameter is updated while keeping the others constant. Such an approach will converge to a local minimum. The update to each parameter is obtained by computing the gradient of the free energy expression given in (\ref{final_free_eq}) w.r.t the parameter and setting it to zero. This leads to three update equations, one each for the mean and covariance matrix of the phase noise and one for updating the mean corresponding to the bits. The update equations are :
\begingroup
\everymath{\small}
\begin{align}
\label{stheta_update_eq}
\bt S_\theta ^T=& \Bigg[\bs\phi_\theta^{-1}+\frac{1}{\sigma^2}\textbf{X}_m\textbf{X}_m^H
+\frac{1}{\sigma^2}\Big ( \text{diag}\left(\textbf{F}^H\textbf{H}\text{diag}(\bs\nu_r +\bs\nu_i-|\textrm{\textit{f}}(\hat{\bt B})|^2)\textbf{H}^H\textbf{F}\right) \Big )\Bigg ] ^{-1} 
\end{align}
\begin{align}
\label{mtheta_update_eq}
 \textbf{m}_\theta=&\textbf{S}_\theta\left[-\frac{1}{\sigma^2}\text{Im}(\textbf{r}^H\textbf{X}_m)^T +\bs\phi^{-1}_\theta\bs\mu_\theta\right]
\end{align}
\begin{align}
\label{real_bit_mean_update_eq}
\bt t_{\hat{b}rk}=&\bt t_{\mu r k}+\frac{1}{\sigma^2}\Bigg[\text{diag}(\bs\alpha_k)\Re\{\bt Z^H\bt r\}-\text{diag}(\bs\delta_k)\bt M_1^T\bt 1 
 + \text{diag}(\bs\alpha_k)\Big(j\Im\{\bt M_0\}\textrm{\textit{f}}_i(\hat{\bt B}_i)-\Re\{\bt M_2^T\}\textrm{\textit{f}}_r({\hat{\bt B}_r})\Big )\Bigg]
\end{align}
\begin{align}
\label{imag_bit_mean_update_eq}
\bt t_{\hat{b}ik}=&\bt t_{\mu i k}+\frac{1}{\sigma^2}\Bigg[ \text{diag}(\bs\beta_k)\Im\{\bt Z^H\bt r\}-\text{diag}(\bs\Omega_k)\bt M_1^T\bt 1 
+ \text{diag}(\bs\beta_k)\Big ( \Im\{\bt M_0\}\textrm{\textit{f}}_r(\hat{\bt B}_r)
+j\Re\{\bt M_2^T\}\textrm{\textit{f}}_i({\hat{\bt B}_i})\Big ) \Bigg]
\end{align}
\endgroup
In the above equations, $\bt t_{\hat{b}rk}$ and $\bt t_{\hat{b}ik}$ represent $\tanh^{-1}(\hat{\bt b}_{rk})$ and $\tanh^{-1}(\hat{\bt b}_{ik})$ (computed element wise) respectively, while $\bt t_{\mu r k}$ and $\bt t_{\mu i k}$ represent $\tanh^{-1}(\bs\mu_{rk})$ and $\tanh^{-1}(\bs\mu_{ik})$ respectively, with $\bs\mu_{rk}$ and $\bs\mu_{ik}$ defined analogous to $\hat{\bt b}_{rk}$ and $\hat{\bt b}_{ik}$ but with prior means. Further, the derivatives $\frac{\partial \textrm{\textit{f}}_r({\hat{\bt B}_r})}{\partial \hat{ \bt b}_{rk}}$ and  $\frac{\partial \textrm{\textit{f}}_i({\hat{\bt B}_i})}{\partial \hat{ \bt b}_{ik}}$ are denoted as $\text{ diag}(\bs\alpha _k)$\hspace{0.1cm} and\hspace{0.1cm} $\text{ diag}(\bs\beta _k)$. The derivatives $\frac{\partial \bs\nu_r}{\partial \hat{\bt b}_{rk}}$ and $\frac{\partial \bs\nu_i}{\partial \hat{\bt b}_{ik}}$ are denoted as $\text{ diag}(\bs\delta_k)$ and $\text{ diag}(\bs\Omega_k)$. Detailed derivation of the update equations is given in Appendix \ref{appendix_free_energy_min}.
\subsection{The Variational Inference Algorithm}

 The pseudo code given in the next page gives the steps involved in the VI based soft bit detection algorithm. It was noted that because of the the coordinate descent approach to free energy minimization, the algorithm converges to a local minimum and not the global minimum. This was particularly found to be an issue in scenarios of high CPE. To overcome this, we use the second term of (\ref{free_energy_eq2}) (denoted as $F_2$ henceforth) as an indicator of convergence to the right PHN sequence. Whenever $F_2$ is greater than a certain threshold, it indicates that the estimated PHN sequence does not correspond to the expected statistics and suggests that the algorithm has converged to a wrong sequence. In such scenarios, we ignore the output of the algorithm and compute the soft bit estimates ignoring PHN.

\begin{algorithm}[t]
\caption{Bit level Variational Inference Algorithm }
\label{alg:bit_level_VI}
\begin{algorithmic}[1] 
\STATE Initialize. $\mu_{nl}\leftarrow 0$ or given priors; $\bs\mu_\theta \leftarrow$ \bt 0; $\bs \phi_\theta \leftarrow \bs \Phi$; $m_{nl} \leftarrow 0$; $\bt m_\theta \leftarrow \bt 0$; $\bt S_\theta \leftarrow \bt 0$;
\STATE Compute $\bt t_{\mu rk}$ and $\bt t_{\mu i k}$ for $k \in \{1,2,\hdots \frac{L}{2}\}$.
\FOR {s=1:$num\_iter$}
\STATE Compute $\bs\nu_r$, $\bs\nu_i$, $\bs\alpha_k$, $\bs \delta_k$, $\bt X_m$.
\STATE Update $\bt S_\theta$ using (\ref{stheta_update_eq}).
\STATE Update $\bt m_\theta$ using (\ref{mtheta_update_eq}).
\FOR{k=L/2:1}
\STATE Compute $\bt t_{\hat{b}rk}$ and $\bt t_{\hat{b}ik}$ using (\ref{real_bit_mean_update_eq}) and (\ref{imag_bit_mean_update_eq}).
\FOR{p=1:L/2}
\STATE Update $\hat{\bt B}$, $\bs\alpha_p$, $\bs\beta_p$, $\bs\delta_p$, $\bs\Omega_p$.
\ENDFOR
\ENDFOR
\ENDFOR
\STATE Compute $F_2$ in (\ref{free_energy_eq2}).
\IF {$F_2 > threshold$}
\STATE Ignore PHN and detect symbols.
\ELSE
\STATE Return
$2(\bt t_{\hat{b}rk}-\bt t_{\mu rk})$ $\&$ $2(\bt t_{\hat{b}ik}-\bt t_{\mu ik})$ for $k \in \{1,2,\hdots \frac{L}{2}\}$.
\ENDIF
\end{algorithmic}
\end{algorithm}

\section{Simulation results }


To test the performance of the suggested algorithm we set up the following simulation. We simulated a link using 64-QAM constellation and OFDM with 64 sub-carriers over a frequency selective channel. The channel was assumed to be a Rayleigh multipath fading channel with 10 taps and an exponential power delay profile. The sampling rate was set to 20 MHz, which in turn implies a  sub-carrier spacing  of 312.5 kHz. The oscillator bandwidth was set to 100 kHz and the standard deviation $\sigma_\theta$ was set to 3$^\circ$. The MATLAB code presented in \cite{Webster} was used to generate the PHN sequences.

\begin{figure}
\centering
\includegraphics[width=5in]{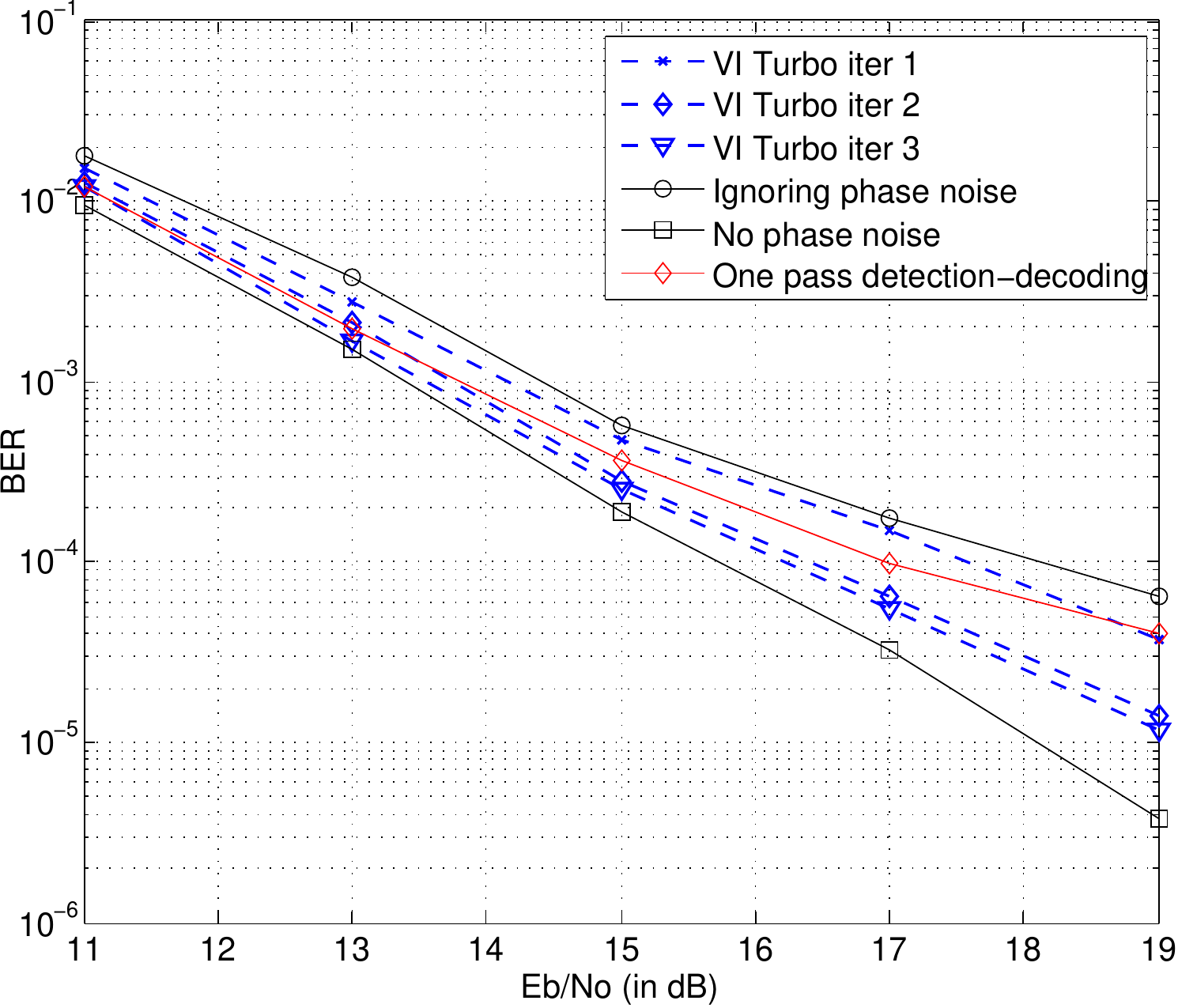}
\caption{BER plot of the turbo receiver.}
\label{fig_2}
\end{figure}


For the turbo receiver setup, an outer LDPC code of rate 3/4 and length 2304 was used. On the transmit side, the message bits were encoded, interleaved and mapped to symbols from the 64-QAM constellation. The length of the outer code was chosen so as to span 6 OFDM symbols. On the receiver side, after the reception of the 6 OFDM symbols, each was passed through the bit level detection algorithm and the extrinsic information so obtained was passed to the soft decoder after deinterleaving. To begin with, the detector was initialized to uniform priors; in subsequent iterations, the extrinsic information obtained from the decoder was used as priors. For this particular simulation, we used 3 outer iterations with 6 iterations for the decoder and 5 iterations for the detector. Fig. \ref{fig_2} presents the results of this simulation and shows a significant improvement in performance with every subsequent iteration of the turbo receiver. It can also be seen that the turbo receiver clearly out-performs a one pass receiver where there is no loop back between the detector and decoder. The receiver performance is compared against scenarios where there is no PHN and where PHN is completely ignored. For both these cases, and for the one pass setup, the FEC decoder was run for 18 iterations.

\section{Conclusion}
In this paper we highlighted the consequences of high CPE and its effect on decision directed algorithms. To overcome this effect, we developed a bit-level soft detection algorithm using the VI approach and used this algorithm to build a turbo receiver. Through simulations we have established the performance gains that can be achieved using the turbo receiver.

\appendices
\section{Evaluation of the free energy expression}
\label{appendix_free_energy_eval}
As shown in (\ref{free_energy_eq2}), the free energy expression can be written as the sum of five terms. Since we have adopted a discrete distribution for $Q_{\bt B}(\bt B)$, it is straightforward to compute the first and third terms. The second and fourth term together constitute the Kullback-Leibler divergence between two multi-variate  Gaussian distributions and can be computed easily. We discuss the computation of the fifth term here. Now,

\begin{align}
\label{free_term_5_eq1}
-2\sigma^2(\romannumeral5 )= &const.+\int_{\textbf{B},\boldsymbol\theta} \Big [ Q(\textbf{B})Q(\boldsymbol\theta)(\textbf{r}-\textbf{PF}^H\textbf{H}\textrm{\textit{f}(\textbf{B})})^H 
(\textbf{r}-\textbf{PF}^H\textbf{H}\textrm{\textit{f}(\textbf{B})}) \Big ] d\textbf{B}d\boldsymbol\theta 
\end{align}
Proceeding exactly as in \cite{darryl1}, we can simplify (\ref{free_term_5_eq1}) to 
\begin{align}
\label{free_term_5_eq2}
-2\sigma^2(\romannumeral5 )
=const.+\int_{\textbf{B}}Q(\textbf{B})\Big[\textrm{\textit{f}(\textbf{B})}^H\textbf{Z}^H\textbf{Z}\textrm{\textit{f}(\textbf{B})}-\textbf{r}^H\textbf{Z}\textrm{\textit{f}(\textbf{B})} 
-\textrm{\textit{f}(\textbf{B})}^H\textbf{Z}^H\textbf{r}+\textrm{\textit{f}(\textbf{B})}^H\boldsymbol\Psi \textrm{\textit{f}(\textbf{B})}\Big]d\textbf{B} 
\end{align}
Here, we have defined $\bt Z=\text{diag}(\bt 1+j\bt m_\theta)\bt F^H\bt H$ and $\bs \Psi=\bt H^H \bt F \text{diag}(\bt S_\theta) \bt F^H \bt H)$. Using Lemmas 5 and 6 from \cite{darryl2}, we can simplify this to 
\begin{align}
-2\sigma^2(\romannumeral5 )=&const.-\textbf{r}^H\textbf{Z}\textrm{\textit{f}($\hat{\bt{B}}$)}-\textrm{\textit{f}($\hat{\bt{B}}$)}^H\textbf{Z}^H\textbf{r}+ \textbf{1}^H(\bt M_1)(\bs{\nu}_r +\bs{\nu}_r) 
+\textrm{\textit{f}$_r$($\hat{\bt{B}}_r$)}^H (\bt M_0)\textrm{\textit{f}$_i(\hat{\bt{B}}_i$)}\nonumber\\   &+\textrm{\textit{f}$_i(\hat{\bt{B}}_i)$}^H (\bt M_0)\textrm{\textit{f}$_r$($\hat{\bt{B}}_r$)} 
+\textrm{\textit{f}$_r$($\hat{\bt{B}}_r$)}^H(\bt M_2)\textrm{\textit{f}$_r$($\hat{\bt{B}}_r$)} 
+\textrm{\textit{f}$_i$($\hat{\bt{B}}_i$)}^H(\bt M_2)\textrm{\textit{f}$_i$($\hat{\bt{B}}_i$)}.
\end{align}
In the equation above, we have used the following definitions. 
\begin{align}
\bt M_0=&(\boldsymbol\Psi+\textbf{Z}^H\textbf{Z}) \\
\bt M_1=&(\boldsymbol\Psi+\textbf{Z}^H\textbf{Z}-\text{diag} (\boldsymbol\Psi+\textbf{Z}^H\textbf{Z}))\\
\bt M_2=&\text{ diag}(\boldsymbol\Psi+\textbf{Z}^H\textbf{Z})  \\
 \bs{\nu}_r=&\sum_{0<i \leq j<(L/2)}2^{i+j} 
\sprod\displaylimits_{p=i}^{p=j}\hat{\bt{b}}_p+\sum_{i=1}^{L/2}2^{2(i-1)}  \\
  \bs{\nu}_i=&\sum_{0 < i \leq j<(L/2)}2^{i+j} 
\sprod\displaylimits_{p=i}^{p=j}\hat{\bt{b}}_{p+(L/2)}+\sum_{i=1}^{L/2}2^{2(i-1)}   
\end{align}

\section{Computing the gradient}
\label{appendix_free_energy_min}
\subsection{Gradient w.r.t $\bt S_\theta$}
Using (\ref{final_free_eq}) and (\ref{free_energy_eq}), we can write
\begin{align}
\frac{\partial \textbf{F}}{\partial\textbf{ S}_\theta^{-1}}=&\frac{\partial (-\romannumeral1-\romannumeral2+\romannumeral3+\romannumeral4-\romannumeral5 )}{\partial \bt{S}_\theta^{-1}} \nonumber  \\
\end{align}

Using results from \cite{matrix_reference}, one can compute the terms above to be : 
\begin{align}
\label{stheta_1to4_eq}
\frac{\partial (\romannumeral1)}{\partial \bt{S}_\theta^{-1}}=0; \hspace{0.75cm}
\frac{\partial (\romannumeral2)}{\partial \bt{S}_\theta^{-1}}=\frac{1}{2}\bt S_\theta^T\bs\phi_\theta^{-1}\bt S_\theta^T ;\hspace{0.75cm}
\frac{\partial (\romannumeral3)}{\partial \bt{S}_\theta^{-1}}=0; \hspace{0.75cm}
\frac{\partial (\romannumeral4)}{\partial \bt{S}_\theta^{-1}}=&\frac{1}{2}\bt S_\theta^T 
\end{align}

To compute $\frac{\partial (\romannumeral5)}{\partial \bt{S}_\theta^{-1}}$, note the following two results \cite{darryl1} :
\begin{align}
\label{stheta_aid_eq1}
\text{(a)\hspace{2.3cm}} \frac{\partial (\bt x^{H}\text{diag}(\bt S_\theta) \bt y)}{\partial \bt{S}_\theta^{-1}}=&\frac{\partial \left(\text{tr}(\text{diag}(\bt x^{H})\bt S_\theta \text{diag}(\bt y))\right)}{\partial \bt{S}_\theta^{-1}}\nonumber \\
=&\frac{\partial \left(\text{tr}(\text{diag}(\bt y)\text{diag}(\bt x^{H})\bt S_\theta )\right)}{\partial \bt{S}_\theta^{-1}} \nonumber \\
=&-\bt S_\theta^T\left(\text{diag}(\bt y)\text{diag}(\bt x^{H})\right) \bt S_\theta \\
\label{stheta_aid_eq2}
\text{(b)\hspace{2cm}}\frac{\partial \left(\text{tr}(\bt X\text{diag}(\bt S_\theta))\right) }{\partial \bt{S}_\theta^{-1}}=&\frac{\partial \left(\text{tr}(\text{diag}(\bt X) \bt S_\theta)\right) }{\partial \bt{S}_\theta^{-1}} \nonumber \\
=&-\bt S_\theta^T\left(\text{diag}(\bt X)\right)\bt S_\theta^T
\end{align}
In the above equations, by diag(\bt S$_\theta$) we mean the diagonal matrix formed using the diagonal entries of the matrix \bt S$_\theta$. Using (\ref{stheta_aid_eq1}) and (\ref{stheta_aid_eq2}) we can compute $\frac{\partial (\romannumeral5)}{\partial \bt{S}_\theta^{-1}}$ to be
\begin{align}
\label{stheta_fifth_eq}
\frac{\partial (\romannumeral5)}{\partial \bt{S}_\theta^{-1}}=&\frac{-1}{2\sigma^2}\Big[ -\bt{S}_\theta^T(\bt{X}_m\bt{X}_m^H)\bt{S}_\theta^T 
-\bt{S}_\theta^T\text{diag}(\bt{F}^H\bt{H}\text{diag}(\bs\nu_r +\bs\nu_i-|\textrm{\textit{f}}(\hat{\bt{B}})|^2)\bt{H}^H\bt{F})\bt{S}_\theta^T \Big]
\end{align}

Using (\ref{stheta_1to4_eq}) and (\ref{stheta_fifth_eq}) it is straightforward to compute (\ref{stheta_update_eq}).

\subsection{Gradient w.r.t $\bt m_\theta$}

Using (\ref{free_energy_eq2}) and (\ref{final_free_eq}, we can write  
\begin{align}
\frac{\partial \textbf{F}}{\partial\textbf{ m}_\theta}=&\frac{\partial (-\romannumeral1-\romannumeral2+\romannumeral3+\romannumeral4-\romannumeral5 )}{\partial \bt{m}_\theta}. 
\end{align}

Using results from \cite{matrix_reference}, one can compute the terms above to be : 
\begin{align}
\label{mtheta_1to4_eq}
\frac{\partial (\romannumeral1)}{\partial \bt{m}_\theta}=0; \hspace{0.75cm}
\frac{\partial (\romannumeral2)}{\partial \bt{m}_\theta}=\bs \phi_\theta^{-T}\bs \mu_\theta-\bs\phi_\theta^{-1}\bt m_\theta; \hspace{0.75cm}
\frac{\partial (\romannumeral3)}{\partial \bt{m}_\theta}=0; \hspace{0.75cm}
\frac{\partial (\romannumeral4)}{\partial \bt{m}_\theta}=0;
\end{align}

To compute $\frac{\partial (\romannumeral5)}{\partial \bt{m}_\theta}$, note the following two results. For any two given vectors \bt x and \bt y, using results from \cite{matrix_reference}, we have :
\begin{enumerate}[(a)]
\item
\begin{align}
\label{mtheta_aid_eq1}
\frac{\partial \left(\bt x^H(\bt I+j\text{diag}(\bt m_\theta))^H(\bt I+j\text{diag}(\bt m_\theta))\bt y \right)}{\partial\bt m_\theta}
=&\frac{\partial \left( (\bt 1+j\bt m_\theta)^H\text{diag}(\bt x^H)\text{diag}(\bt y)(\bt 1+j\bt m_\theta) \right)}{\partial\bt m_\theta}\nonumber \\
=&-2\text{diag}(\bt x^H)\text{diag}(\bt y)\bt  m_\theta
\end{align}
\item
\begin{align}
\label{mtheta_aid_eq2}
\frac{\partial\left(\bt x ^H \text{diag}(\bt Z^H\bt Z)\bt y \right)}{\partial \bt m_\theta} 
=&\frac{\partial \left(\text{tr}(\text{diag}(\bt x^H)\bt Z^H\bt Z\text{diag}(\bt y))\right)}{\partial \bt m_\theta} \nonumber \\
=&\frac{\partial \left((\bt 1 +j\bt m_\theta)^T \text{diag}(\bt F^H \bt H \text{diag}(\bt y)\text{diag}(\bt x^H)\bt H^H \bt F)
(\bt 1 -j\bt m_\theta)\right)} {\partial \bt m_\theta} \nonumber \\
=&2\text{diag}(\bt F^H \bt H \text{diag}(\bt y)\text{diag}(\bt x^H)\bt H^H \bt F)\bt m_\theta
\end{align}
\end{enumerate}

Using (\ref{mtheta_aid_eq1}) and (\ref{mtheta_aid_eq2}), we can compute $\frac{\partial (\romannumeral5) }{\partial \bt m_\theta}$ to be :
\begin{align}
\label{mtheta_fifth_eq}
 \frac{\partial (\romannumeral5) }{\partial \bt m_\theta}=&\frac{-1}{\sigma^2}\Bigg[\bt X_m^H\bt X_m\bt m_\theta + \Im (\bt r^H \bt X_m)^T 
+ \text{diag}(\bt F^H \bt H \text{diag}\Big (\nu_r +\nu_i-|\textrm{\textit{f}}(\hat{\bt B})|^2) \Big)\bt m_\theta  \Bigg]
\end{align}

Using (\ref{mtheta_1to4_eq}), (\ref{mtheta_fifth_eq}) and the update equation for $\bt S_\theta$ given in (\ref{stheta_update_eq}), we can compute the update equation for $\bt m_\theta$ to be as given in (\ref{mtheta_update_eq}).

\subsection{Gradient w.r.t $\hat{\bt b}_{rk}$ and $\hat{\bt b}_{ik}$ }

To compute \Large $\frac{\partial F}{\partial \hat{\bt b}_{rk}}$,
\normalsize consider the term (\textit{i}) from (\ref{free_energy_eq2}):
\begin{equation}
\label{term_1}
(\romannumeral1)=\sum _{n=1}^N \sum _{l=1}^{L} \left[ \left(\frac{1+\hat{b}_{nl}}{2} \right)\log \left(\frac{1+\mu_{nl}}{2} \right)+ \left(\frac{1-\hat{b}_{nl}}{2} \right)\log \left(\frac{1-\mu_{nl}}{2} \right) \right]
\end{equation}

Differentiating (\ref{term_1}) w.r.t to $\hat{b}_{nl}$ gives $\tanh^{-1}(\mu_{nl}).$ Thus, if we denote  $\frac{\partial(\romannumeral1)}{\partial \hat{\bt b}_{rk}}$ as $\bt t_{\mu rk}$ and $\frac{\partial(\romannumeral1)}{\partial \hat{\bt b}_{ik}}$ as $\bt t_{\mu ik}$ for $k \in \{1,2,\hdots \frac{L}{2}\}$, we have $\bt t_{\mu rk}=\tanh^{-1}$($\bs \mu_{rk}$) and $\bt t_{\mu ik}=\tanh^{-1}$($\bs\mu_{ik}$). Similarly, differentiating term (\textit{ii}) (given in (\ref{24})) w.r.t to $\hat{b}_{nl}$ gives $\tanh^{-1}(\hat{b}_{nl})$. Denoting  $\frac{\partial(\romannumeral2)}{\partial \hat{\bt b}_{rk}}$ as $\bt t_{\hat{b}rk}$ and $\frac{\partial(\romannumeral2)}{\partial \hat{\bt b}_{ik}}$ as $\bt t_{\hat{b}ik}$, we have $\bt t_{\hat{ b} rk}=\tanh^{-1}$($\hat{\bt b}_{rk}$) and $\bt t_{\hat{ b} ik}=\tanh^{-1}$($\hat{\bt b}_{ik}$).

\normalsize
\begin{equation}
\label{24}
(\romannumeral2)=\sum _{n=1}^N \sum _{l=1}^{L} \left[ \left(\frac{1+\hat{b}_{nl}}{2} \right)\log \left(\frac{1+\hat{b}_{nl}}{2} \right)+ \left(\frac{1-\hat{b}_{nl}}{2} \right)\log \left(\frac{1-\hat{b}_{nl}}{2} \right) \right]
\end{equation}

Clearly, the terms ($\romannumeral3$) and ($\romannumeral4$) of \ref{free_energy_eq2} are independent of $\hat{\bt b}_{rk}$ and $\hat{\bt b}_{ik}$. It remains to compute the derivative of the fifth term. To compute $\frac{\partial(\romannumeral5)}{\partial \hat{\bt b}_{rk}}$ and $\frac{\partial(\romannumeral5)}{\partial \hat{\bt b}_{ik}}$, we note the following :
\begin{enumerate}[(a)]
\item From \cite{darryl2},
\begin{align}
\label{grad_b_eq1}
 \frac{\partial \textrm{\textit{f}}_r({\hat{\bt B}_r})}{\partial \hat{ \bt b}_{rk}}=&\text{ diag}(\bs\alpha _k)  \hspace{0.24cm}
\text{where,  } \bs\alpha_k=& \text{I}\big(k=L/2\big)2^{k-1}.\bt 1+\sum_{l=1}^{k}2^{l-1}\sprod\displaylimits _{p=l,p\neq k}^{L/2} \hat{\bt b}_{rp}
\end{align}
\item From \cite{darryl2},
\begin{align}
\label{grad_b_eq2}
 \frac{\partial \textrm{\textit{f}}_i({\hat{\bt B}_i})}{\partial \hat{\bt b}_{ik}}=&\text{ diag}(\bs\beta _k)   \hspace{.24cm}
\text{where,  }\bs\beta_k=& j.\Big(\text{I}\big(k=L/2\big)2^{k-1}.\bt 1+\sum_{l=1}^{k}2^{l-1}\sprod\displaylimits _{p=l,p\neq k}^{L/2} \hat{\bt b}_{ip}\Big)
\end{align}
\item  For any given matrix \bt M with complex entries,
\begin{align}
\label{grad_b_eq3}
 \frac{\partial\textrm{\textit{f}}_r({\hat{\bt B}_r})^H \bt M \textrm{\textit{f}}_r({\hat{\bt B}_r}) }{\partial\hat{ \bt b}_{rk}}=\frac{\partial\textrm{\textit{f}}_r({\hat{\bt B}_r})^H \bt M \textrm{\textit{f}}_r({\hat{\bt B}_r})}{\partial \textrm{\textit{f}}_r({\hat{\bt B}_r})}\frac{\partial \textrm{\textit{f}}_r({\hat{\bt B}_r})}{\partial \hat{\bt b}_{rk}} 
=2 \Big [\Re \{\textrm{\textit{f}}_r({\hat{\bt B}_r})^H \bt M\}\text{diag}(\bs \alpha_{rk}) \Big ]^T
\end{align}
\item For any given matrix \bt M with complex entries, 
\begin{align}
\label{grad_b_eq4}
 \frac{\partial\textrm{\textit{f}}_i({\hat{\bt B}_i})^H \bt M \textrm{\textit{f}}_i({\hat{\bt B}_i}) }{\partial \hat{\bt b}_{ik}}=\frac{\partial\textrm{\textit{f}}_i({\hat{\bt B}_i})^H \bt M \textrm{\textit{f}}_i({\hat{\bt B}_i})}{\partial \textrm{\textit{f}}_i({\hat{\bt B}_i})}\frac{\partial \textrm{\textit{f}}_i({\hat{\bt B}_i})}{\partial\hat{ \bt b}_{ik}} 
=2 \Big [\Im \{\textrm{\textit{f}}_i({\hat{\bt B}_i})^H \bt M\}\text{diag}(\bs \beta_{ik}) \Big ]^T
\end{align}
\item From \cite{darryl2},
\begin{align}
\label{grad_b_eq5}
 \frac{\partial \bs\nu_r}{\partial \hat{\bt b}_{rk}}=\hspace{0.1cm}\text{diag}\Bigg(\text{I} \big(0<k<L/2\big).2^{2k}\bt 1 +\sum_{\substack{ 0<i\leq k \leq j < L/2 \\ i\neq j}}\sprod\displaylimits_{\substack{ p=i \\ p \neq k}}^{j}\hat{b}_{rp}\Bigg )
\triangleq \hspace{0.1cm}\text{diag}( \bs \delta_{k})
\end{align}
\item From \cite{darryl2},
\begin{align}
\label{grad_b_eq6}
 \frac{\partial \bs\nu_i}{\partial \hat{\bt b}_{ik}}=&\hspace{0.1cm}\text{diag}\Bigg(\text{I}\big(0<k<L/2\big).2^{2k}\bt 1 +\sum_{\substack{ 0<i\leq k \leq j < L/2 \\ i\neq j}}\sprod\displaylimits_{\substack{ p=i \\ p\neq k}}^{j}\hat{b}_{ip}\Bigg ) \triangleq\hspace{0.1cm}\text{diag}(\bs\Omega_{k})
\end{align}
\item For any given matrix \bt M,
\begin{equation}
\label{grad_b_eq7}
\frac{\partial (\bt 1^T \bt M \bs\nu_r)}{\partial \hat{\bt b}_{rk}}=\bt M^T\bt 1\text{diag}(\bs\delta_k)
\end{equation}
\item For any given matrix \bt M,
\begin{equation}
\label{grad_b_eq8}
\frac{\partial (\bt 1^T \bt M \bs\nu_i)}{\partial \hat{\bt b}_{ik}}=\bt M^T\bt 1\text{diag}(\bs\Omega_k)
\end{equation}
\end{enumerate}

In (\ref{grad_b_eq5}) and (\ref{grad_b_eq6}), I($\cdot$) represents the indicator function which is 1 if the argument is 0 and 0 otherwise. Using all of the results from equations (\ref{grad_b_eq1}) to (\ref{grad_b_eq8}), we can compute the gradient to be :
\begin{align}
\text{\hspace{-0.14cm}}\frac{\partial F}{\partial \hat{\bt b}_{rk}}&=\bt t_{\mu r k}-\bt t_{\hat{b}rk}+\frac{1}{\sigma^2}\Bigg[ \text{diag}(\bs\alpha_k)\left (\Re\{\bt Z^H\bt r\}-\Re\{\bt M_2^T\}\textrm{\textit{f}}_r({\hat{\bt B}_r}) \right )\nonumber \\
&\hspace{3cm}+j\text{diag}(\bs\alpha_k)\Im\{\bt M_0\}\textrm{\textit{f}}_i(\hat{\bt B}_i)-\text{diag}(\bs\delta_k)\bt M_1^T\bt 1\Bigg]
\end{align}
\begin{align}
\frac{\partial F}{\partial \hat{\bt b}_{ik}}&=\bt t_{\mu i k}-\bt t_{\hat{b}ik}+\frac{1}{\sigma^2}\Bigg[ \text{diag}(\bs\beta_k)\left( \Im\{\bt Z^H\bt r\}+ \Im\{\bt M_0\}\textrm{\textit{f}}_r(\hat{\bt B}_r)\right )\nonumber \\
&\hspace{3cm}+j.\text{diag}(\bs\beta_k)\Re\{\bt M_2^T\}\textrm{\textit{f}}_i({\hat{\bt B}_i})-\text{diag}(\bs\Omega_k)\bt M_1^T\bt 1\Bigg]
\end{align}





\bibliographystyle{IEEEtran}
\bibliography{ref_file_long}

\begin{thebibliography}{10}
\providecommand{\url}[1]{#1}
\csname url@samestyle\endcsname
\providecommand{\newblock}{\relax}
\providecommand{\bibinfo}[2]{#2}
\providecommand{\BIBentrySTDinterwordspacing}{\spaceskip=0pt\relax}
\providecommand{\BIBentryALTinterwordstretchfactor}{4}
\providecommand{\BIBentryALTinterwordspacing}{\spaceskip=\fontdimen2\font plus
\BIBentryALTinterwordstretchfactor\fontdimen3\font minus
  \fontdimen4\font\relax}
\providecommand{\BIBforeignlanguage}[2]{{%
\expandafter\ifx\csname l@#1\endcsname\relax
\typeout{** WARNING: IEEEtran.bst: No hyphenation pattern has been}%
\typeout{** loaded for the language `#1'. Using the pattern for}%
\typeout{** the default language instead.}%
\else
\language=\csname l@#1\endcsname
\fi
#2}}
\providecommand{\BIBdecl}{\relax}
\BIBdecl

\bibitem{petrovic2}
D.~Petrovic, W.~Rave, and G.~Fettweis, ``Performance degradation of
  coded-{OFDM} due to phase noise,'' in \emph{Vehicular Technology Conference,
  2003. VTC 2003-Spring. The 57th IEEE Semiannual}, vol.~2, April 2003, pp.
  1168--1172 vol.2.

\bibitem{Webster}
\BIBentryALTinterwordspacing
M.~Webster and M.~Seals. Suggested phase noise model for 802.11hrb. [Online].
  Available:
  \url{https://mentor.ieee.org/802.11/dcn/00/11-00-0296-01-00sb-suggested-phas%
e-noise-model-for-802-11-hrb.ppt}
\BIBentrySTDinterwordspacing

\bibitem{pollet1}
T.~Pollet, M.~Van~Bladel, and M.~Moeneclaey, ``{BER} sensitivity of {OFDM}
  systems to carrier frequency offset and {W}iener phase noise,''
  \emph{Communications, {IEEE} Transactions on}, vol.~43, no. 234, pp.
  191--193, Feb/Mar/Apr 1995.

\bibitem{petrovic1}
D.~Petrovic, W.~Rave, and G.~Fettweis, ``Phase noise influence on bit error
  rate, cut-off rate and capacity of {M-QAM} {OFDM} signaling,'' in \emph{In
  Proc. Intl. OFDM Workshop (InOWo)02}, 2002.

\bibitem{wu1}
S.~Wu and Y.~Bar-Ness, ``{OFDM} systems in the presence of phase noise:
  consequences and solutions,'' \emph{Communications, IEEE Transactions on},
  vol.~52, no.~11, pp. 1988--1996, Nov. 2004.

\bibitem{wu2}
------, ``Performance analysis on the effect of phase noise in {OFDM}
  systems,'' in \emph{Spread Spectrum Techniques and Applications, 2002 IEEE
  Seventh International Symposium on}, vol.~1, 2002, pp. 133--138 vol.1.

\bibitem{darryl1}
D.~D. Lin and T.~J. Lim, ``The variational inference approach to joint data
  detection and phase noise estimation in {OFDM},'' \emph{Signal Processing,
  {IEEE} Transactions on}, vol.~55, no.~5, pp. 1862--1874, May 2007.

\bibitem{septier1}
F.~Septier, Y.~Delignon, A.~Menhaj-Rivenq, and C.~Garnier, ``Monte {C}arlo
  methods for channel, phase noise, and frequency offset estimation with
  unknown noise variances in {OFDM} systems,'' \emph{Signal Processing, IEEE
  Transactions on}, vol.~56, no.~8, pp. 3613--3626, Aug. 2008.

\bibitem{gong1}
Y.~Gong and X.~Hong, ``A new algorithm for {OFDM} joint data detection and
  phase noise cancellation,'' in \emph{Communications, 2008. {ICC} '08. IEEE
  International Conference on}, May 2008, pp. 636--640.

\bibitem{rave1}
W.~Rave, D.~Petrovic, and G.~Fettweis, ``Iterative correction of phase noise in
  multicarrier modulation,'' in \emph{In Proc. of the 9th International {OFDM}
  Workshop {InOWo}}, 2004.

\bibitem{barbieri1}
A.~Barbieri, G.~Colavolpe, and G.~Caire, ``Joint iterative detection and
  decoding in the presence of phase noise and frequency offset,''
  \emph{Communications, IEEE Transactions on}, vol.~55, no.~1, pp. 171--179,
  Jan. 2007.

\bibitem{darryl3}
D.~D. Lin, R.~Pacheco, T.~J. Lim, and D.~Hatzinakos, ``Joint estimation of
  channel response, frequency offset, and phase noise in {OFDM},'' \emph{Signal
  Processing, IEEE Transactions on}, vol.~54, no.~9, pp. 3542--3554, Sept.
  2006.

\bibitem{darryl2}
D.~D. Lin and T.~J. Lim, ``Bit-level equalization and soft detection for
  {G}ray-coded multilevel modulation,'' \emph{Information Theory, {IEEE}
  Transactions on}, vol.~54, no.~10, pp. 4731--4742, Oct. 2008.

\bibitem{matrix_reference}
\BIBentryALTinterwordspacing
M.~Brookes. Matrix reference manual. [Online]. Available:
  \url{http://www.ee.ic.ac.uk/hp/staff/dmb/matrix/intro.html#Intro}
\BIBentrySTDinterwordspacing

\end{thebibliography}
%

\end{document}